\documentclass[runningheads]{llncs}

\usepackage{amsmath}
\usepackage{amssymb}
\usepackage{graphicx}

\begin{document}

\title{An FPGA-based Solution for Convolution Operation Acceleration}

\titlerunning{FPGA-based Convolution Operation}

\author{Trung Pham-Dinh, Bao Bach-Gia, Lam Luu-Trinh, Minh Nguyen-Dinh, Hai Pham-Duc, Khoa Bui-Anh, Xuan-Quang Nguyen, Cuong Pham-Quoc}

%\author{For blind review requirement}
%
\authorrunning{Trung Pham et al.}

\institute{Ho Chi Minh City University of Technology (HCMUT) \and
Vietnam National University - Ho Chi Minh City (VNU-HCM)
\email{\{trung.pham.ktmt,bao.bachbbace12,lam.luu1602,minh.nguyen207bk,\\hai.phamcse,khoa.bui140,nxquang,cuongpham\}@hcmut.edu.vn}}

\maketitle

\begin{abstract}
Hardware-based acceleration is an extensive attempt to facilitate many computationally-intensive mathematics operations. This paper proposes an FPGA-based architecture to accelerate the convolution operation - a complex and expensive computing step that appears in many Convolutional Neural Network models. We target the design to the standard convolution operation, intending to launch the product as an edge-AI solution. The project's purpose is to produce an FPGA IP core that can process a convolutional layer at a time. System developers can deploy the IP core with various FPGA families by using Verilog HDL as the primary design language for the architecture. The experimental results show that our single computing core synthesized on a simple edge computing FPGA board can offer 0.224 GOPS. When the board is fully utilized, 4.48 GOPS can be achieved.

\end{abstract}



\section{Introduction}\label{sec:introduction}

The prominence of Convolutional Neural Networks (CNN) has sparked computer vision and AI technology. As the demand for these applications grows, scientists have proposed many CNN models and fitted them to many aspects of the digital world. Soon enough, embedded devices find themselves the urge to incorporate the AI capability into their functionality to ease the exhausted computation on the server. However, micro-controllers have limitations regarding the resources needed to perform AI computation. Therefore, FPGAs (Field Programmable Gate Arrays) dominate the battlefield, along with ASICs (Application Specific Integrated Circuits)~\cite{demicheli2013hardware}. The FPGA technology has been rapidly growing, with each new generation integrating more Intellectual Properties (IP) to expand the chip's capability. These IP cores are the engineers' resources to integrate any CNN functionality and inferences to the embedded devices.

This paper proposes our initial attempt to bring AI functionality to edge devices. Our work focuses on convolution operation acceleration using FPGA technology. Convolution is long known for its intensive and time-consuming computing process, and it occupies around 90\% in nearly every CNN model. Therefore, the solution to this acceleration problem will undoubtedly lay the ground for further AI-at-the-edge developments.

The rest of the paper is organized as follows. The convolution operation and related work are presented in Section~\ref{sec:convo}. We describe the overview of the proposed architecture in Section \ref{sec:architecture}. Section \ref{sec:implementation} and \ref{sec:result} will continue to discuss the implementation of the IP core and its behavior under simulation environment, along with its synthesis report on different Xilinx's FPGA families. Finally, Section~\ref{sec:conclusion} concludes the paper.

\section{Background and related work}
\label{sec:convo}
In this section, we introduce an overview of the convolution operation that we are targeting as our hardware-based IP core. We then present related work for the FPGA-based CNN computing cores.

\subsection{Convolution Operation}
Convolution is an operation performing on two arguments referred as \textit{input} (images) and \textit{filters}. The output of the operation is known as the \textit{feature map}. Consider our input image as a two-dimension matrix. The convolution operation is performed as shown in Equation \ref{eq:convo_1}. We apply the element-wise multiplication of the Kernel matrix and a partial matrix of the input Image and take the sum of all the products. We continue to apply the same computation by "sliding" and "weighted-summing" the Kernel on the Image until we yield all the outputs of the Feature Map.
    \begin{equation}
        F(i,j) = (I\circledast K)(i,j) = \sum\limits_{m}\sum\limits_{n}I(i+m,j+n)\times K(m,n)\label{eq:convo_1}
    \end{equation}
    
    % \begin{figure}[h]
    %     \centering
    %     \includegraphics[width=0.7\textwidth]{figures/convo_visualization.png}
    %     \caption{Visualization of the Convolution Operation for $F[0][0]$}
    %     \label{fig:convo_visual}
    % \end{figure}
    
In practical usage, most images come in RGB format, encoded to three matrices corresponding to each dimension of the picture - Red, Green, and Blue. In this case, the Kernel expands to a 3D tensor. Equation \ref{eq:convo_2} describes the resulting Feature Map retrieved by performing the Convolution Operation subjecting to this change.
    \begin{equation}
        F(i,j) = (I\circledast K)(i,j,d) = \sum\limits_{d}\sum\limits_{m}\sum\limits_{n}I(i+m,j+n,d)\times K(m,n,d)\label{eq:convo_2}
    \end{equation}
    
    % \begin{figure}[h]
    %     \centering
    %     \includegraphics[width = 0.9\textwidth]{figures/multi_channel.png}
    %     \caption{Visualization of the Multi-Channel Convolution Operation}
    %     \label{fig:multi_channel}
    % \end{figure}
    
\subsection{Related work}
In recent years, many studies in the literature have been proposed to accelerate CNN in FPGA. Some of them try to improve parallelism by exploiting the huge amount of hardware resources in FPGA such as research in ~\cite{KaiyuanGuo}, ~\cite{HuiminLi2016}, ~\cite{LinXinhan2018}, \cite{ZhiqiangLiu2016}. Another approach is proposals that reduce the complexity of CNN by using lightweight models like BNN such as ~\cite{Yang2019}, ~\cite{Ghasemzadeh2018}, ~\cite{JiaoLi2017}, ~\cite{MossDuncan2017}, ~\cite{NakaharaHiroki2017}, ~\cite{Nurvitadhi2016}, ~\cite{ProstBoucle2017}. However, all these systems focus on high-performance FPGA computing platforms. Meanwhile, in this work, we target edge computing platforms with less amount of hardware resources and low energy consumption.

\section{IP Core Architecture}\label{sec:architecture}
\begin{figure}[h]
    \centering
    \includegraphics[width = 0.6\textwidth]{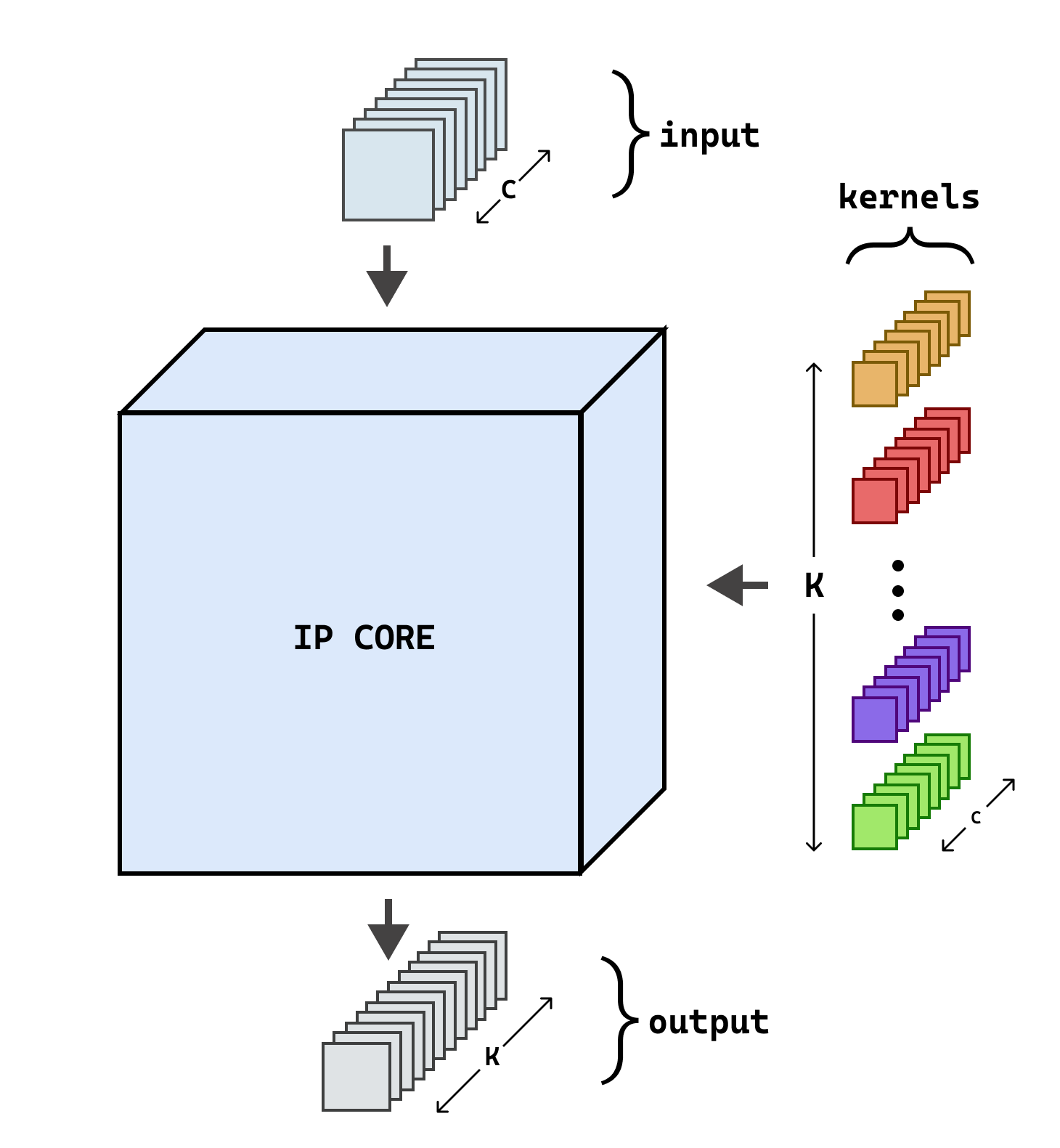}
    \caption{IP Core Functionality}
    \label{fig:ip-core}
\end{figure}

Figure \ref{fig:ip-core} conveys the primary functionality of our IP core. Because the core takes on one layer at a time, it always expects a set of feature maps containing \texttt{C} channels and the associated sets of  \texttt{K} kernels (each kernel includes \texttt{C} channels) as inputs. After the accelerated computation, the IP core produces another set of feature maps, including \texttt{K} channels. These operations sum up the overall functionality of the architecture.

The proposed solution in this paper exploits the profound advantages of SoC architectural design. The integrated on-chip Processing System (PS) will efficiently transfer the input data down the stream. The IP initially loads the two sets of input data into its internal groups of BRAMs (Section \ref{subsec:bram} will further elaborate on the organizing of data among the RAM pool). Since the amount of data is typically large, we use a direct memory access controller, or DMA, to handle the transfer; hence cutting down the workload on the PS.

\begin{figure}[h]
    \centering
    \includegraphics[width = 0.6\textwidth]{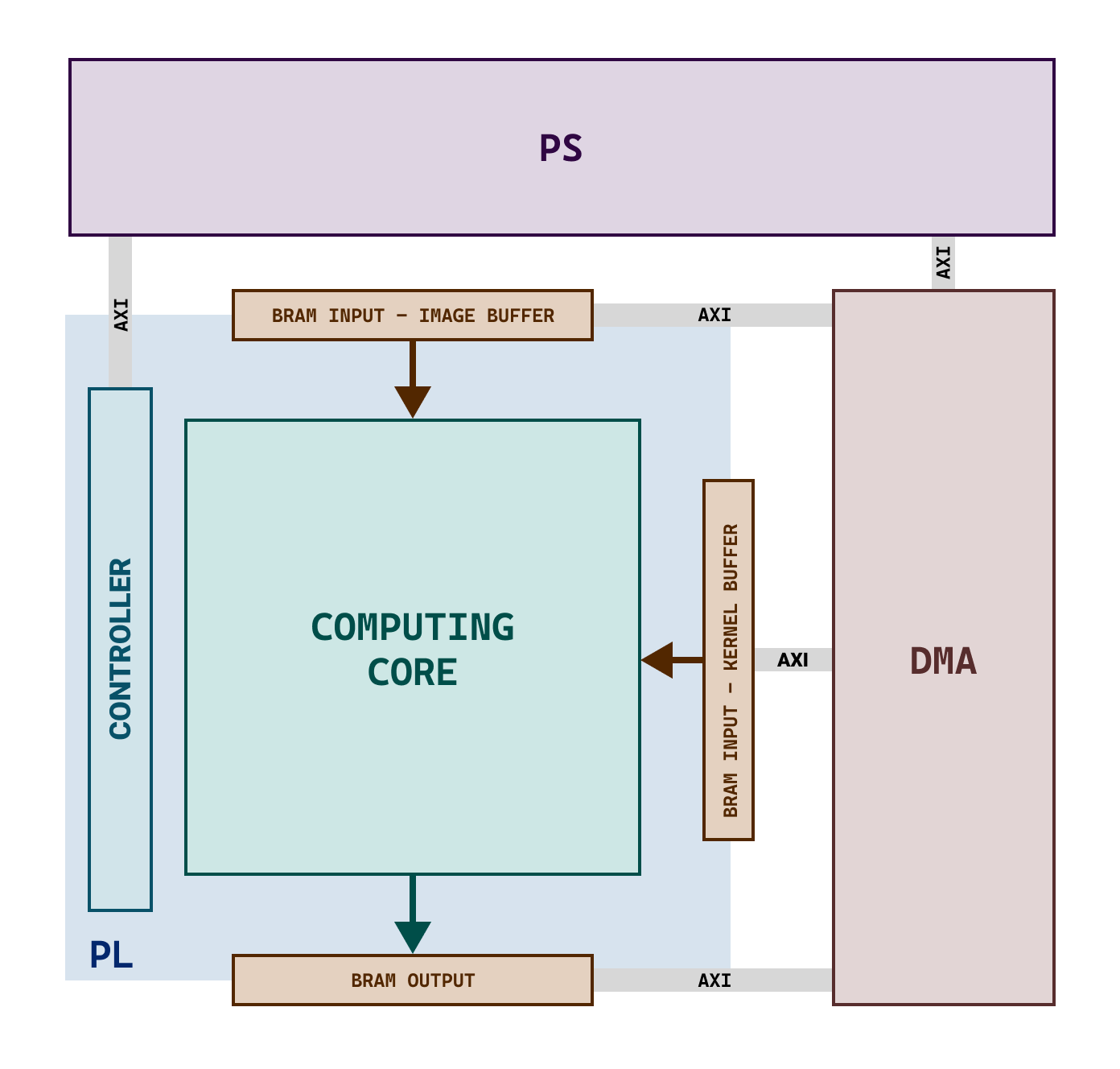}
    \caption{Dataflow}
    \label{fig:dataflow}
\end{figure}
    
The IP has two groups of BRAMs to store the input image and the input kernel separately. The IP core's internal logic enables the Computing Core module when the DMA has finished transferring the input data. In the most abstract sense, the Computing Core is a convolutional computation unit of the architecture. When fed with an image channel and its associated kernels, it spits out the resulting feature map after applying the convolutional operation. The detailed implementation to achieve this functionality is far more complicated and is discussed in Section \ref{subsec:computingcore}.
    
After computing the feature map, the Computing Core will store it in another set of BRAMs. The DMA is continually responsible for transferring the calculated results back to the PS from this set of BRAM. All the communications between the DMA and the BRAMs mentioned so far are through {AXI4} interfaces - a communication interface protocol that Xilinx has adopted as the standard for their IPs communication.
    
Figure \ref{fig:dataflow} illustrates the dataflow discussed so far. The final sub-module to mention here is the Controller unit. To perform a correct convolution operation, it will receive the information needed from the PS (for example, the dimension of the input image and the input kernel). Section \ref{subsec:computingcore} will discuss the Computing core in depth.

\section{Implementation}\label{sec:implementation}
This section presents in detail the architecture of the IP core. We first start describing our memory management. Then, the introduction of the IP's central computing core, including sub-components, will be followed.

\subsection{Memory Management}\label{subsec:bram}
Block RAMs (BRAMs) in FPGA are common resources for storing data as they provide random access to memory elements. Xilinx provides IP Block Memory Generator (BMG) to help designers access BRAM efficiently under various configurations. However, BMG has only two ports for concurrently reading and writing; therefore, the architecture will distribute and organize data into multiple BMGs to exploit this concurrency behavior.

\subsubsection{Input BRAMs}
\begin{figure}[h]
        \centering
        \includegraphics[width=1.0\textwidth]{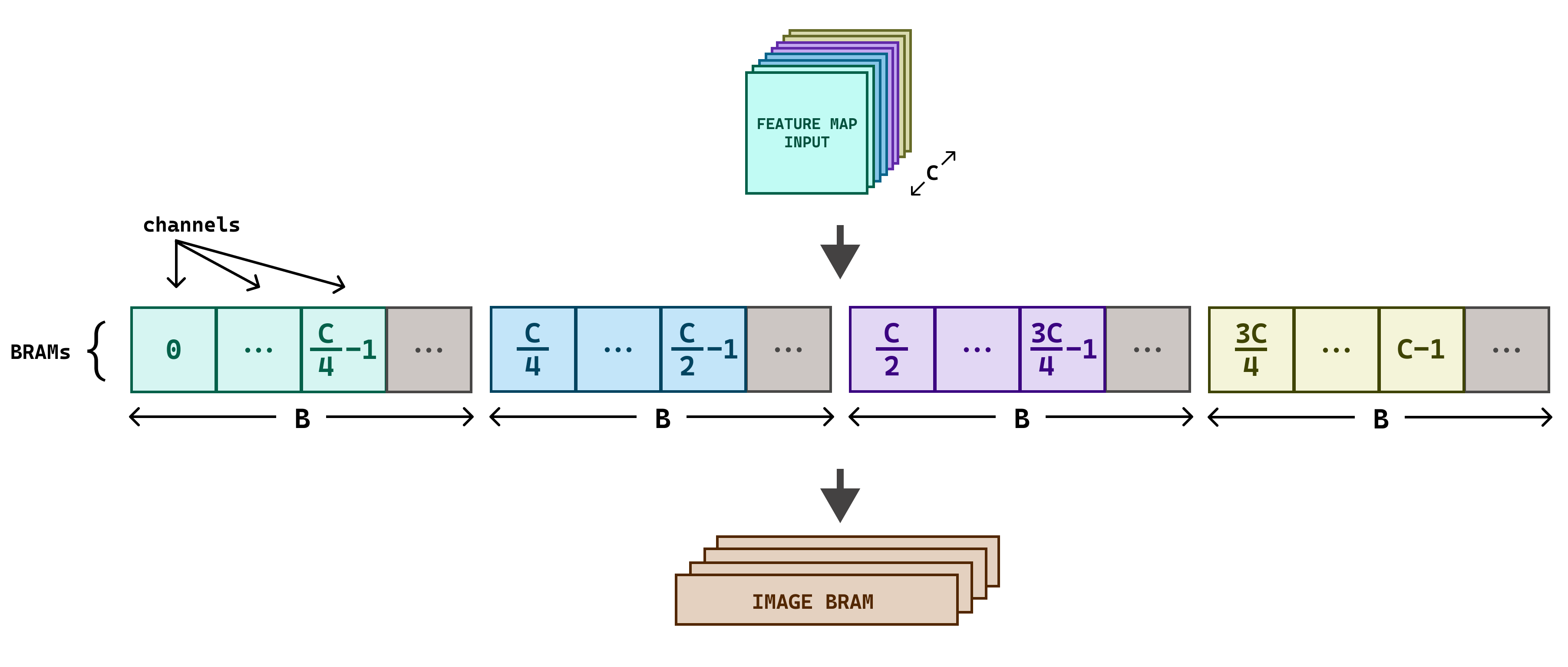}
        \caption{The set of Image BRAMs. Here, $B$ is the largest possible feature map size divided by 4, and $C$ is the total number of channels of this layer's input image (or feature map).}
        \label{fig:image-bram}
    \end{figure}

The architecture has a set of four BMGs for each input image (Figure \ref{fig:image-bram}). Each BMG in the set will store one fourth of the image channels. Because the depth (the number of channels) per image varies, the size of each BMG must be large enough to hold the largest possible image. This also means that for small images, there will be redundant, or meaningless, slots in the BMGs. Here we use 4 BMGs but not some other arbitrary number because the majority of CNN inferences like the {AlexNet} \cite{NIPS2012_c399862d} or the {MobileNet} \cite{howard2017mobilenets} architecture has one common characteristic. For these inferences, all the the produced feature maps are divisible by 4, except for the first input image. By separating the channels into four sets, we can process them in parallel, hence speeding up the computation of the operation.
    
The same divisible-by-4 idea applies to the kernel inputs. However, one standard convolutional step typically requires more than one multiple-channel kernel. Therefore, we organize the kernel data according to the number of kernels. Furthermore, because the number of channels is a multiple of 4, the number of kernels associated with each convolution layer must also be divisible by 4. Therefore, for each BMG that contains one-fourth of the channels of every kernel, we further divide it into four smaller BMGs, each holding one-fourth of the whole kernels. This distribution scheme guarantees that each BMG portion of the image has an associated set of four BMGs, breaking the computing process down to four separate tasks independent of each other.

\subsubsection{Output BRAMs}
Since an output feature map of one convolutional layer is the required input for the next one in a typical CNN model, the distribution of the computed feature map among the output BRAMs is identical to that of the input image BRAMs. It means that each one-fourth of the output channels will be stored in a separate BGM. These output BGMs will then be used to either read the result out (through the DMA) or compute the next convolution layer in the network.

\subsection{Computing Core}\label{subsec:computingcore}
Computing Core is the essential module of the proposed IP core architecture. As stated before, this module is responsible for producing the weighted sum (or partial sum) when feeding the input image and weight data from the BRAMs. Subsequent sections discuss the design and behavior of this module from the multi-channel view to the multi-kernel execution.

\subsubsection{Multi-Channel Architecture}

    \begin{figure}[h]
        \centering
        \includegraphics[width=1.0\textwidth]{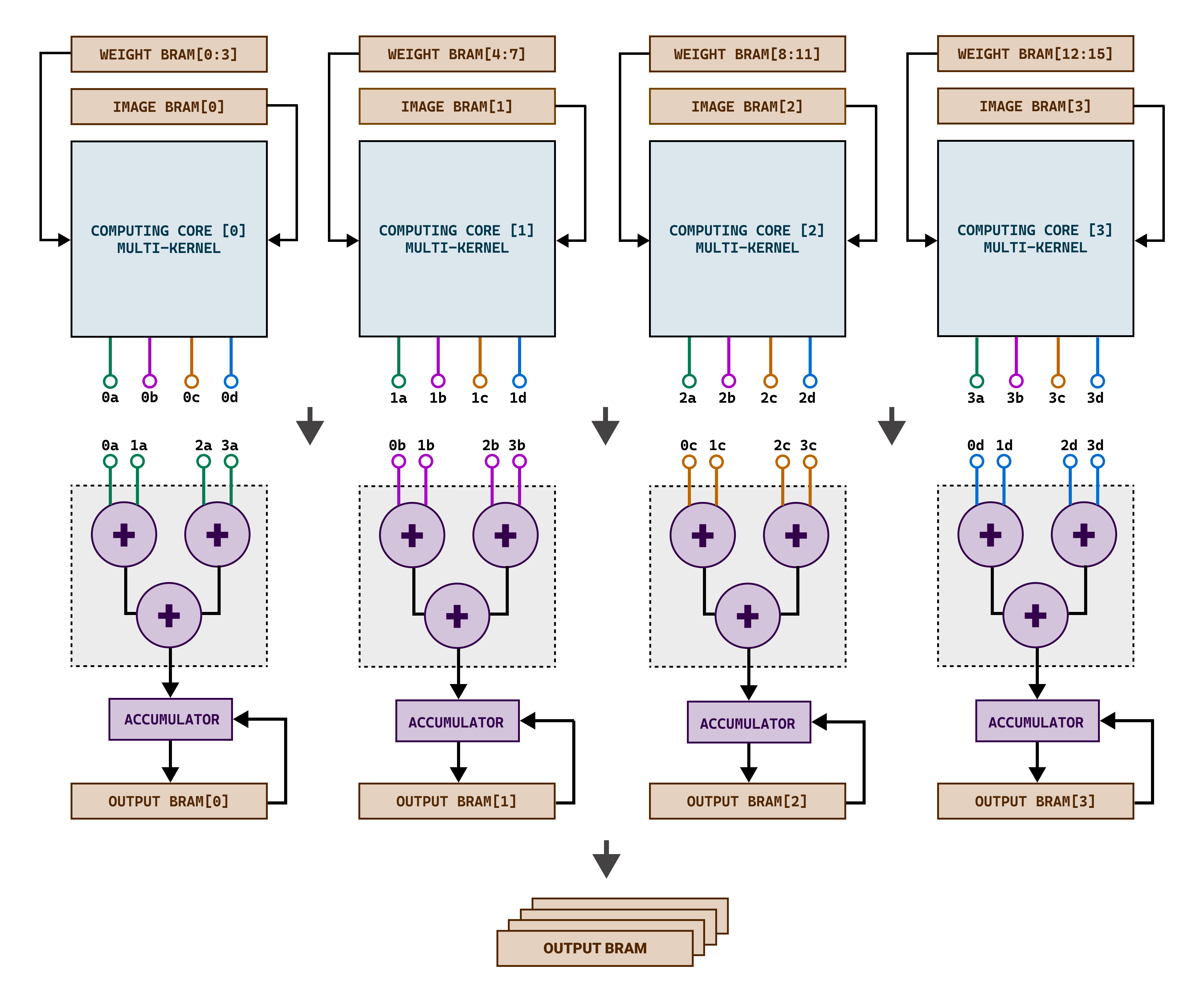}
        \caption{Computing Cores Diagram}
        \label{fig:computing-core-set}
    \end{figure}

The proposed architecture comprises four computing cores; each performs the convolution operation on one-fourth of the depth of the input feature maps. With the organization of the input data among the BRAMs discussed previously, the four computing cores can execute the operation totally in parallel, called multi-channel architecture. Each computing core has one associated image BRAM and four corresponding weight BRAMs - complying exactly with the number of BRAMs in the architecture. Figure \ref{fig:computing-core-set} illustrates the proposed design. In this architecture, computed PSUM values of each core get accumulated continually into the output BRAMs until the processing depth of images is finished. The computing cores will continue to repeat the process but with another set of kernels. It keeps processing until all the kernels have been filtered.

\subsubsection{Multi-Kernel Computing Core}
\begin{figure}[h]
    \centering
    \includegraphics[width=1.0\textwidth]{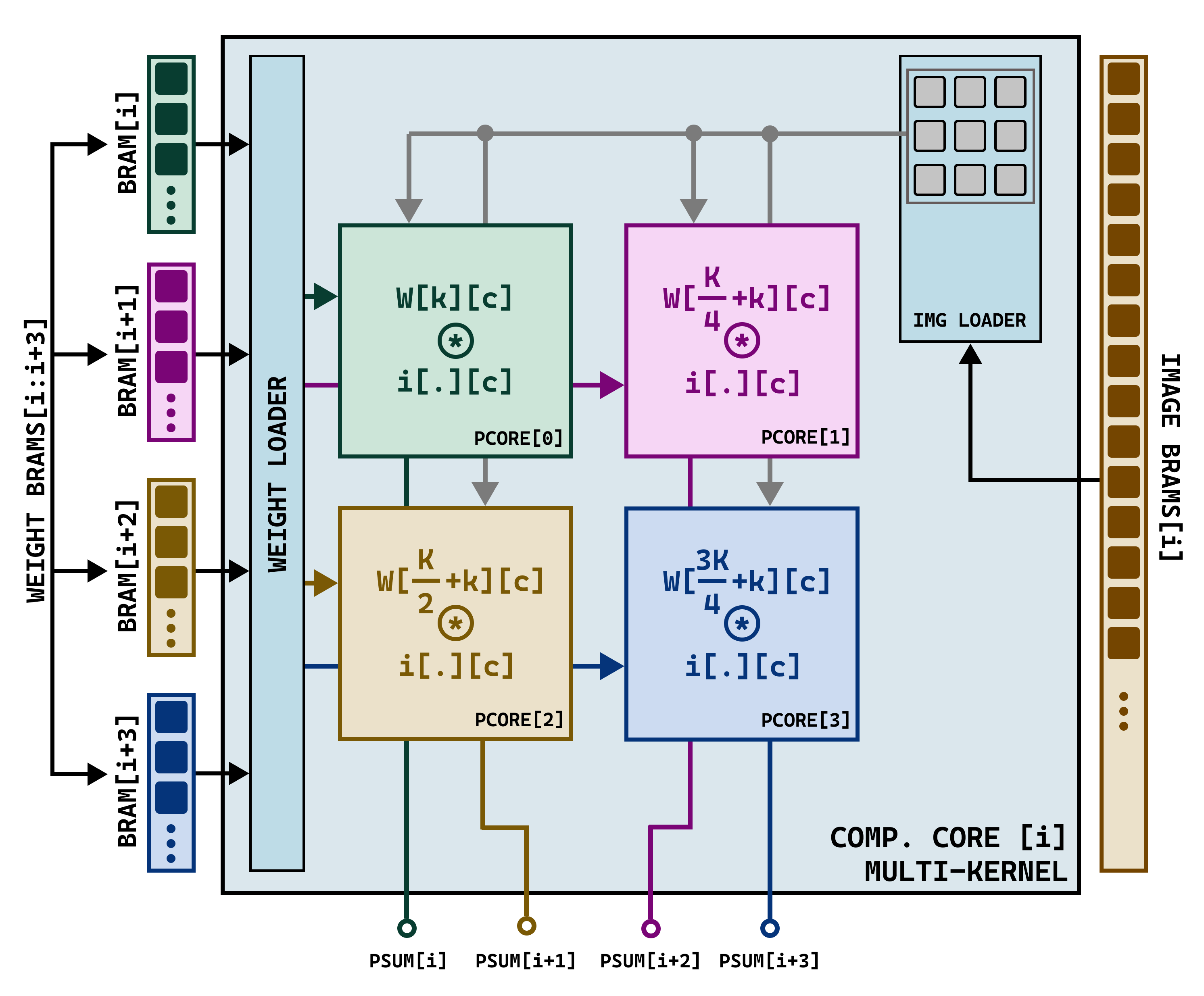}
    \caption{A Single Computing Core. In the diagram, $W[k][c]$ is the channel $c^{th}$ of kernel $k^{th}$, $i[.][c]$ is the partial input image at the channel $c^{th}$, $\circledast$ is the convolution operator, $comp. core[i]$ is the $i^{th}$ computing core. In general, $k$ iterates from $0$ to $K/4 - 1$, where $K$ is the total number of input kernels; and $c$ iterates from $i*C/4$ to $(i+1)*C/4 - 1$, for each iteration of $k$, where $C$ is total number of input kernels.}
    \label{fig:computing-core}
\end{figure}
    Each computing core in the architecture produces four PSUM (Partial Sums) values due to four kernels received, as shown in Figure  \ref{fig:computing-core}. The produced PSUM values belong to the one-fourth output feature map channel, speeding up the computing process four times more. A single computing core contains four more sub-module called PCOREs. The transfer of data from the BRAMs to the PCOREs is through the intermediate loaders. Each PCORE computes a PSUM value according to the weight input it receives from the Weight Loader, while the Image Loader holds a set of nine pieces of input values for all the four PCOREs. This computing model is weight stationery since the weight data persists in the weight loader.
    
In contrast, the image loader continually fetches different input images after each computed set of PSUMs. The internal logic of a PCORE is simple. Each contains a set of MAC (Multiply and Accumulate) units and adder modules to perform a weighted-sum operation.

\subsubsection{Pipeline}
Essentially, the computing process of the computing cores can be divided into two stages. The first stage transfers sufficient data from the BRAMs to the loaders while the second stage performs the convolution operation and accumulates the computed PSUMs into the output BRAMs. The architecture pipelines these two stages to accelerate the computation further, effectively cutting down the wasted cycles.

\subsubsection{Bias Handling}
We utilize a handful feature from Xilinx's BMG IP to handle the bias input. The input bias is first to get initialized into the output BRAMs through the PS. Since the computed PSUMs from the computing core get accumulated into these BRAMs, it effectively causes the same effect as if the biases were added to the weighted sum. Therefore there is no logic needed to handle the bias.

\section{Experimental Results}\label{sec:result}

In this section, we present our synthesis results with three different Xilinx FPGA devices, including xc7z020clg400, xc7z020clg484, and xzcu3eg-sbva484-1. We then anlyze simulation results to claim the throughput of our computing core.

\subsection{Synthesis Results}\label{subsec:simulation-result}
Table \ref{table:utilization2} show the resources used on different FPGAs including the number of Look-up Table (LUTs), Flip-flop (FF), and percentage of these resources. Maximum working frequencies for operation are calculated based on "Data Path Delay" values of the synthesis. According to the table, our computing core consumes less than 5\% hardware resources of the Pynq Z2 board, one of the most suitable FPGA boards for edge computing. In other words, we can deploy up to 20 cores concurrently to further improve computing performance.

%\begin{table}[]
%\centering
%\begin{tabular}{l||l|l|l|}
%\hline
%\textbf{Resource} & \textbf{Utilization} & \textbf{Available} & \textbf{Utilization \%} \\ \hline
%LUT               & 5027                 & 53200              & 9.45                    \\ \hline
%LUTRAM            & 124                  & 17400              & 0.71                    \\ \hline
%FF                & 4959                 & 106400             & 4.66                    \\ \hline
%BUFG              & 1                    & 32                 & 3.13                    \\ \hline
%\end{tabular}
%\caption{Synthesis report table of using resources on Pynq-Z2}
%\label{table:utilization}
%\end{table} 

%\begin{figure}[h]
%    \centering
%    \includegraphics[width = 1\textwidth]{figures/utilization_pynq.png}
%    \caption{Synthesis report of using resources on Pynq-Z2 from Vivado}
%    \label{fig:utilization}
%\end{figure}

\begin{table}[]
\centering
\caption{Synthesis result on different FPGAs}
\begin{tabular}{l||r|r|r}
\hline
\textbf{FPGA}       & \textbf{\#LUTs} & \textbf{\#FF} & \textbf{Max frequency} \\ \hline\hline
xc7z020clg400-1     & 5027   (9.45\%)         & 4959  (4.66\%)        & 112 MHz                \\ \hline
xc7z020clg484-1     & 5243   (9.86\%)         & 5054  (4.75\%)        & 93 MHz                 \\ \hline
xzcu3eg-sbva484-1-i & 11917  (16.89\%)        & 14522 (10.29\%)       & 161 MHz                \\ \hline
\end{tabular}
\label{table:utilization2}
\end{table}

\subsection{Simulation Results}\label{subsec:synthesis-result}

\begin{figure}[h]
    \centering
    \includegraphics[width = 1\textwidth]{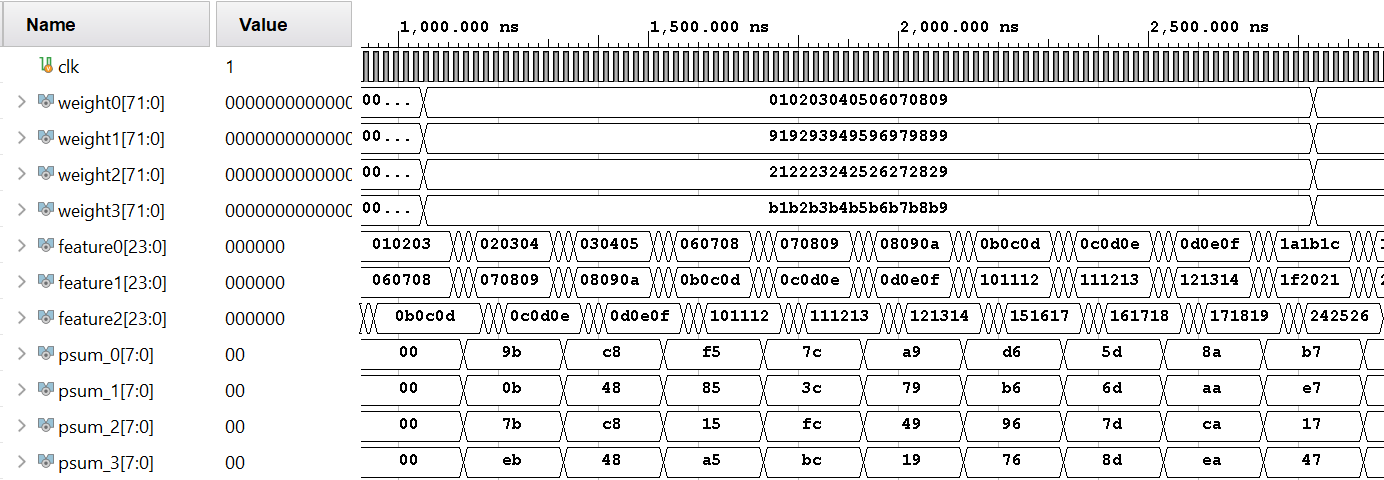}
    \caption{One part of the waveform from the simulation of a single Computing core}
    \label{fig:waveform}
\end{figure}

Figure \ref{fig:waveform} describes a part of the simulation waveform of one Computing core on Vivado. Each feature signal (feature 0, feature1, feature2) concatenates three 8-bit feature entries to create a 3x3 tile of the feature. Each weight signal (weight0, weight1, weight2, weight3) links nine 8-bit weight entries, so each weight signal represents a channel of input weight. Because each Computing core can compute convolution on four different kernels simultaneously, four weight signals correspond to four different kernels. As a result, there are four partial sum signals (psum0, psum1, psum2, psum3), and each signal results from a single weight channel overlapping over a single 3x3 tile of the feature. The computing core needs eight clock cycles to compute four psum values and accumulate them to BRAM outputs. The system has four Computing cores to compute sixteen psum values for each eight clock cycles.

% total psum = no_kernel*no_channel*(feature_width - 2)^2 %
%time = (total psum) / (psum/clock_cycle) / freq %
The system consists of four Computing cores; a single core can compute four psum values for each eight clock cycles. With the input feature of [224x224x8] and input weight of [8x3x3x8], the system needs to compute 3,154,176 psum values. Based on Synthesis Result, the IP core can operate on 112MHz clock frequency with the Pynq Z2 FPGA board. Therefore, we can deduce the theory time needed for computing this sample, which is 0.01408 seconds. In other words, the throughput of a single core is 0.224 GOPS. With the Pynq Z2 FPGA board for edge computing devices mentioned above, when 20 cores are deployed, our computing system can offer up to 4.48 GOPS.

\section{Conclusion}\label{sec:conclusion}

With the rapid growth in AI, the demand of real time processing is significantly high. Convolution operation is the core of convolution neural network and take up most of the computation time. Therefore, this paper proposes a FPGA-based architecture for accelerating convolution operation. The architecture is implemented by Verilog and be able to operate on different FPGA devices. This architecture can compute on multiple channels of feature and multiple kernels of filter simultaneously, which increases the parallelism on the computation. In addition, load stage and computation stage are pipelined, which significantly reduces the computation time and reachs 0.224 GOPS on a single core. This core is synthesized on different FPGAs and reachs maximum frequency of 161MHz with reasonable used resources. Future optimization on this architecture is try to make it more flexible and able to operate on a higher frequency to prove the great prospect of FPGA in AI processing.

\section*{Acknowledgement}
This research is funded by Vietnam National University - Ho Chi Minh City (VNU-HCM) under grant number B2021-20-02. We acknowledge the support of time and facilities from Ho Chi Minh City University of Technology (HCMUT), VNU-HCM for this study.

\bibliographystyle{splncs04}
\bibliography{citation}
%\nocite{*}  % generate reference without citing

\end{document}